\documentclass[a4paper]{jpconf}
\usepackage{graphicx}
\usepackage[compatibility=false]{caption}
\usepackage{subcaption}
\usepackage{empheq} 
\usepackage{siunitx} 

\usepackage{comment}
\usepackage[colorlinks=true, allcolors=blue]{hyperref} 
\usepackage{color, soul}

\newcommand{\db}{drive bunch }
\newcommand{\p}{proton }%
\newcommand{\wrt}{with respect to }

\begin{document}
\title{Predicting the Trajectory of a Relativistic Electron Beam for External Injection in Plasma Wakefields}

\author{F. Pe{\~n}a Asmus$^{1,2}$, F. M. Velotti$^3$, M. Turner$^3$, S. Gessner$^3$, M. Martyanov$^1$, C. Bracco$^3$, B. Goddard$^3$, P. Muggli$^1$}

\address{$^1$Max Planck Institute for Physics, Munich, Germany}
\address{$^2$Technical University of Munich, Munich, Germany}
\address{$^3$CERN, Geneva, Switzerland}

\ead{pena@mpp.mpg.de}

\begin{abstract}
We use beam position measurements over the first part of the AWAKE electron beamline, together with beamline modeling, to deduce the beam average momentum and to predict the beam position in the second part of the beamline. %
Results show that using only the first five beam position monitors leads to much larger differences between predicted and measured positions at the last two monitors than when using the first eight beam position monitors. %
These last two positions can in principle be used with ballistic calculations to predict the parameters of closest approach of the electron bunch with the proton beam. %
In external injection experiments of the electron bunch into plasma wakefields driven by the proton bunch, only the first five beam position monitors measurements remain un-affected by the presence of the much higher charge proton bunch. %
Results with eight beam position monitors show the prediction method works in principle to determine electron and proton beams closest approach within the wakefields width ($<$1\,mm), corresponding to injection of electrons into the wakefields. %
Using five beam position monitors is not sufficient. %
\end{abstract}

\section{Introduction}

External injection of electrons in wakefields driven by a self-modulating charged particle bunch~\cite{bib:kumar} is challenging. %
In the case of a positively charged drive bunch (e.g., protons), the wakefields driven by the un-modulated \db are predominantly defocusing for electrons along the entrance plasma density ramp (if any) and over the beginning of the plasma. %
In addition, as the self-modulation (SM) process develops, the phase velocity of the wakefields is slower than that of the \db \cite{bib:pukhov, bib:schroeder}. %
It can be so slow that even low energy electrons (a few tens of MeV) dephase with respect to the wakefields. %
Electrons can thus be lost through defocusing by transverse wakefields. %

In a continuous plasma longer than the growth length of the SM process to allow for acceleration, external injection must occur after the SM process has saturated. %
Oblique injection (mrad angle) a few meters ($\sim$5) into the plasma is in principle necessary~\cite{bib:caldwell}. %

The electron beam, which is the wake-field acceleration witness, is transported to the plasma cell via a newly built~\cite{bib:Chiara_challenge,bib:Schmidt_primary_beamline} and commissioned~\cite{bib:Chiara_systematic} transfer line.

In the Advanced Wakefield Experiment (AWAKE) case~\cite{bib:muggli}, the \p bunch population (1 to $3\times10^{11}$) vastly exceeds that of the electron bunch ($\sim10^9$). %
The \p bunch signal overwhelms beam position monitors (BPMs) we use to monitor beam trajectories, without intercepting the beams along the common beamline, just before the plasma entrance. %
These BPMs do not allow for simultaneous measurements of the two beams trajectories during injection events. %
Screens cannot be used because they would grossly scatter the low energy incoming electron bunch (19\,MeV). %
They would also be damaged by, and stop the laser pulse we use for plasma creation and self-modulation seeding. 
We therefore tested a method to predict the electron beam trajectory from BPM measurements made in the upstream beamline that is separate from the proton line, and that is thus not affected by the \p bunch presence. %
The goal is to predict the position of the electron bunch at the last two BPMs where the beams propagate ballistically towards the plasma (no magnets). %
These predictions can then be used to find the point of closest approach (distance $d$ between the electron and proton beam and location $z$) from these two positions (ballistic propagation). %

\begin{figure}[!htbp]
    \centering
    \includegraphics[width=\linewidth]{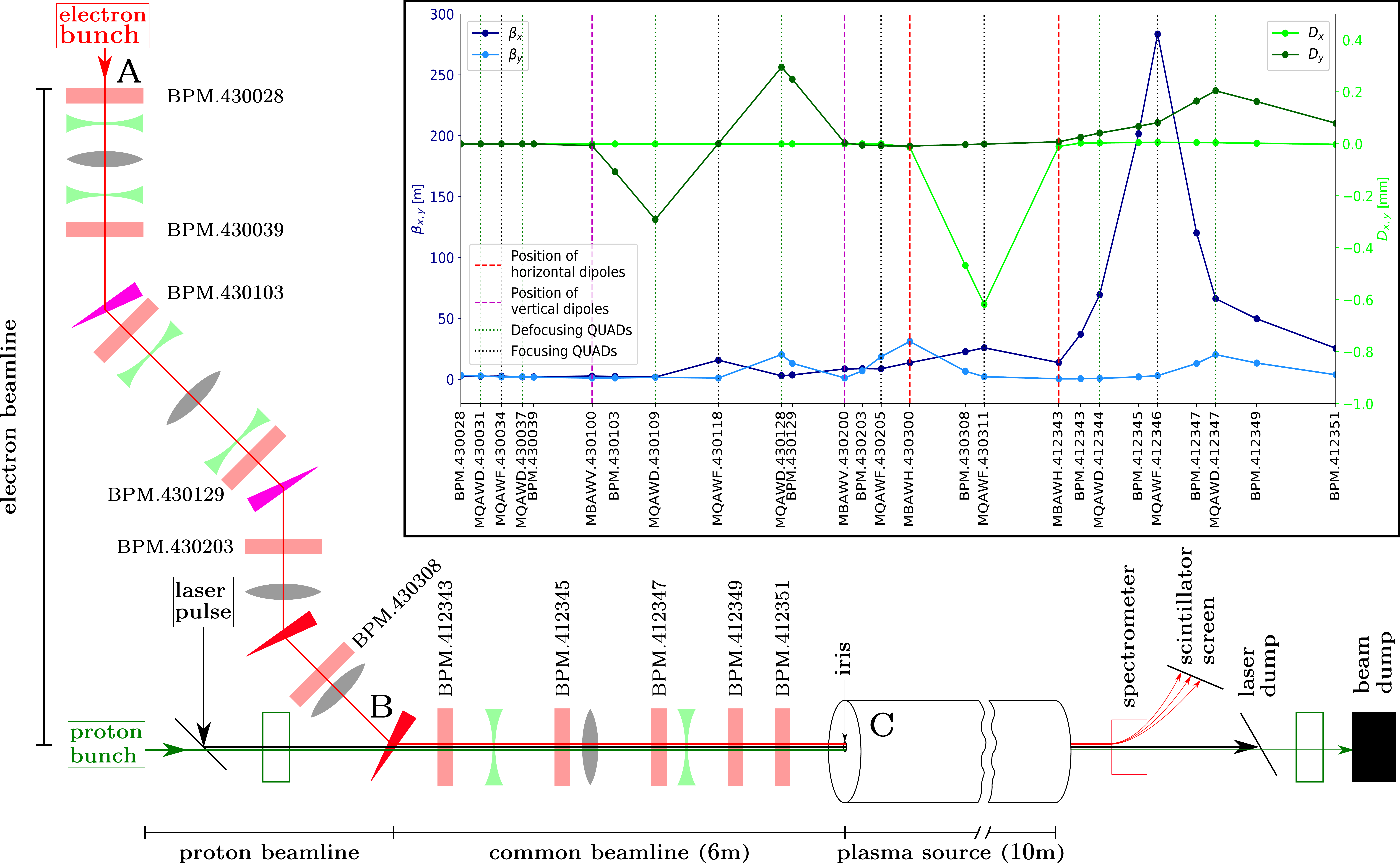}
    \caption{\label{fig:AWAKE_facility}
    Schematic of the electron beamline designed to transport the beam from the electron source to the plasma entrance. %
    Red rectangles show BPMs, purple/red triangles dipole magnets bending in the y-/x-plane, green shapes de-focusing and grey shapes focusing (in x-) magnetic quadrupoles. %
    The electron only beamline is located between points A and B and the electron/proton common line between points B and C. %
    The laser pulse creating the plasma propagates simultaneously with the proton and electron bunches through the common beamline and the plasma source. %
    Inset: %
    Amplitude function $\beta$ (x-plane dark blue, y-plane light blue line) and dispersion $D$ (x-plane light green, y-plane dark green line) along the beamline (s). %
    Beam normalized emittance 2\,mm-mrad. %
    Quadrupole magnets are labelled as MQAW followed by F for focusing and D for defocusing. %
    Dipole magnets are labelled by MBAW followed by H for horizontal and V for vertical bending, respectively. %
    Note the vertical achromatic dog-leg between dipole magnets MBAWV.430100 and 200 and the horizontal achromatic bend between dipole magnets MBAWH.430300 and 412343.}
\end{figure}

A schematic of the AWAKE electron and common electron/proton beamlines is shown in Figure~\ref{fig:AWAKE_facility}. %

\section{Model}

We assume that the transverse position of the electron bunch \wrt the design trajectory (for a given particle momentum) along the beamline (here as measured by $\text{BPM}_i$, with $i=1,...,9$ at position $s_i$) is determined by the dispersion and betatron contributions (valid for both transverse planes labeled with $x$ and $y$): %

\begin{equation} \label{eq:Main_Assumption}
    y_{{BPM}} (s_i) = \sqrt{\beta(s_i)} \, y_\beta (s_i) + D_y (s_i) \, \delta_p
\end{equation}%

The first term on the right hand side is the betatron contribution, which is the product of the square root of the amplitude or beta function $\beta$ and the betatron oscillation amplitude $y_\beta$. %
The betatron oscillation amplitude describes the oscillation in transverse position of the electron bunch around the design trajectory due to incoming conditions (misaligned position or angle) combined with (de)focusing elements in the beamline and is given by~\cite{bib:Book_Edwards}: %

\begin{equation} \label{eq:betatron_oscillation_theory}
    y_\beta (s_i) = R \cdot \cos(2 \pi \mu(s_i) + \phi),
\end{equation}%

with the Courant-Snyder Invariant $R^2$, the phase advance $\mu$ and a phase shift $\phi$. %

The second term on the right hand side of eq.~\ref{eq:Main_Assumption} is the dispersion contribution, which is the product of the dispersion function $D$ and the momentum offset $\delta_p$. %
The momentum offset is the momentum fraction by which the mean momentum of the electron bunch $p$ differs from the design momentum $p_0$ and is defined by $\delta_p = (p-p_0) / p_0$. %

We obtain the amplitude function $\beta$, the dispersion function $D$ and the phase advance $\mu$ with the Methodical Accelerator Design X (MADX) program~\cite{bib:MADX}. %

In the experiment the electron source delivers a beam that is supposed to follow the reference trajectory along the beamline. %
However, variations in incoming position and angle lead to betatron oscillations along the beamline. %
Variation in input momentum lead to dispersion offsets that are enhanced by the strong chromaticity of the beamline~\cite{bib:Chiara_systematic}. %
We assume that, on average, the beam follows the reference trajectory and therefore compute that trajectory from the average of all measured ones. %
We subtract this reference trajectory from all measurements. %
The results in this paper are shown relative to this reference trajectory (zero displacement). %

\section{Dispersion contribution}

The beamline is designed with two BPMs (430103 and 430129) located at positions separated by approximately $\pi$ in phase, i.e. $\mu_2-\mu_1=0.48$ in eq.~\ref{eq:betatron_oscillation_theory}. %
We can thus use the periodicity of the betatron oscillation to determine $\delta_p$. %
We first assume that $\mu_2 - \mu_1 = 0.5$. %
In this case, the betatron oscillation amplitude at both positions has the same value, but with opposite signs, i.e. $y_{\beta_1} = - y_{\beta_2}$. %
Summing equation~\ref{eq:Main_Assumption} at these two BPM locations and solving for the momentum offset gives %

\begin{equation} \label{eq:Momentum_Offset}
    \delta_p
    = \frac{\sqrt{\beta_1} \cdot y_{BPM_2} + \sqrt{\beta_2} \cdot y_{BPM_1}}
    {\sqrt{\beta_1} \cdot D_{y_2} + \sqrt{\beta_2} \cdot D_{y_1}},
\end{equation}

where the dispersion $D$ and amplitude $\beta$ functions are given by MADX and the positions at the BPMs $y_{BPM}$ are measured. %
This feature of the transfer line is also used in routine operation to calculate the mean incoming beam momentum offset with respect to the line~\cite{bib:Chiara_systematic}.

\begin{figure}[!htpb]
    \begin{subfigure}[t]{0.49\textwidth}
        \includegraphics[width=\linewidth]{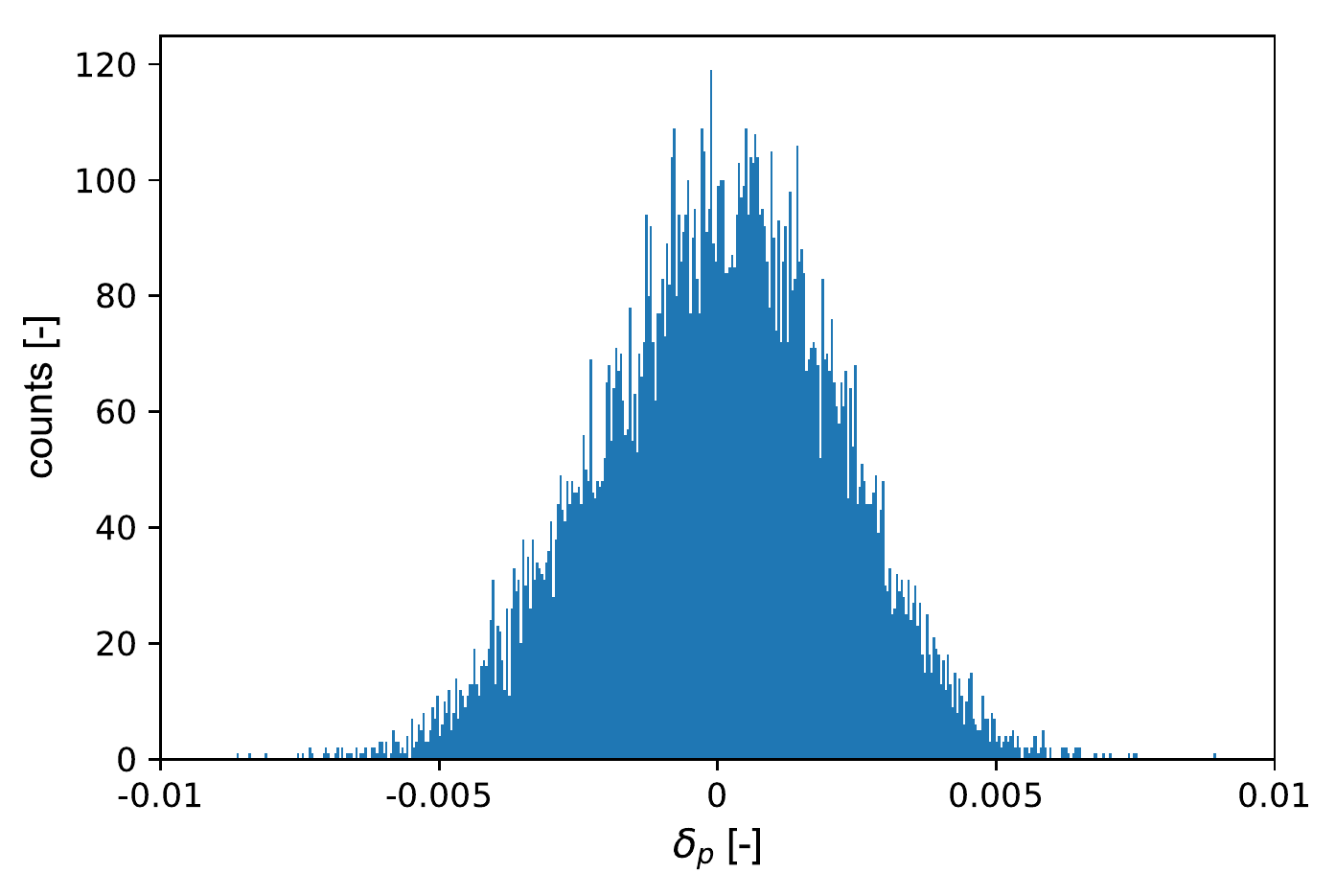}%
        \vspace*{-1.5cm}\caption{\hspace*{5.3cm}\label{subfig:momentum_offset_hist}}
    \end{subfigure}%
    \begin{subfigure}[t]{0.49\textwidth}
        \includegraphics[width=\linewidth]{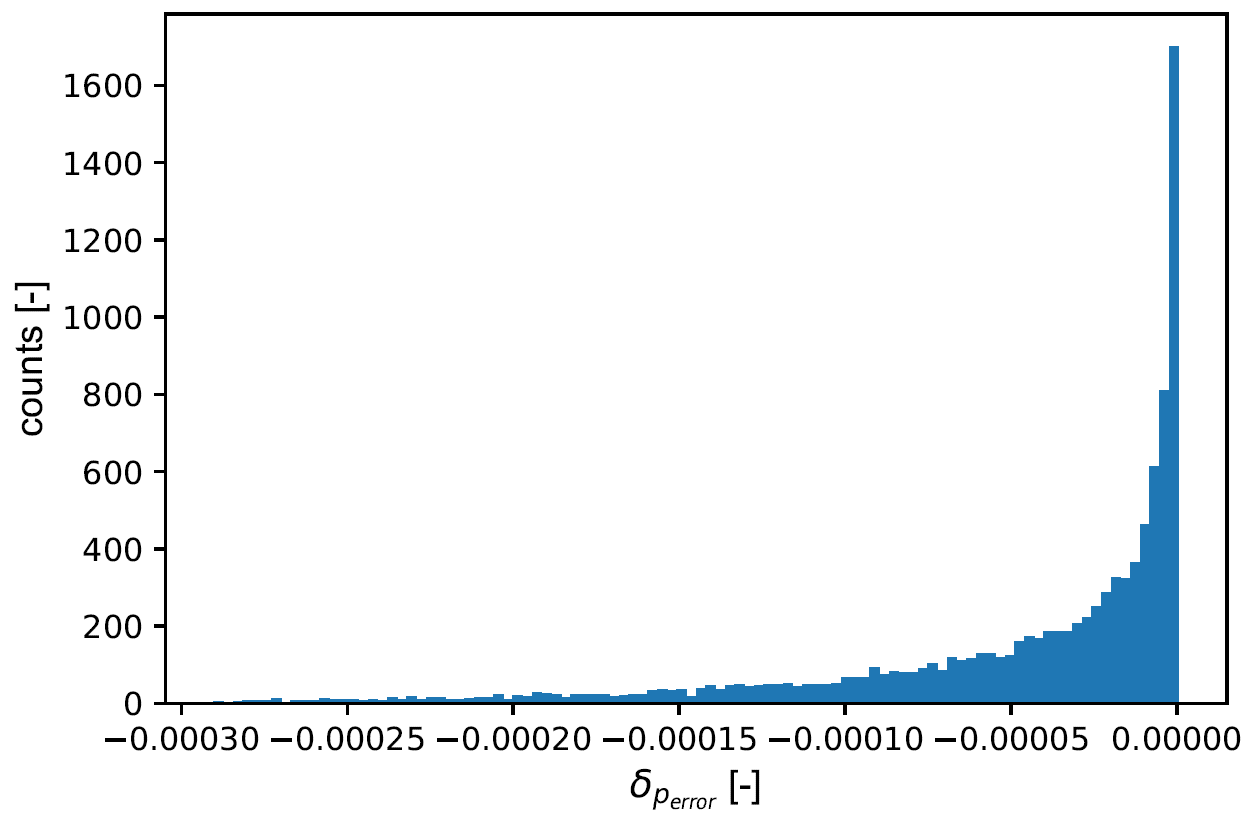}%
        \vspace*{-1.5cm}\caption{\hspace*{5cm}\label{subfig:momentum_offset_error}}
    \end{subfigure}
    \caption{\label{fig:momentum_offset_histograms}(a) Histogram of momentum offset with the assumption $\mu_2 - \mu_1 = 0.5$. (b) Histogram of difference in momenta calculated with assumptions $\mu_2 - \mu_1 = 0.5$ and $\mu_2 - \mu_1 = 0.48.$}
\end{figure}

Figure~\ref{subfig:momentum_offset_hist} shows the histogram of $\delta_p$ determined from $\num{E4}$ consecutive measurements. %
The histogram is centered around zero (by construction) and has an RMS (square root of the variance) of \num{2.1E-3}. %

To quantify the influence of the $\mu_2 - \mu_1 = 0.5$ approximation, we use MADX with the values of $\delta_p$ obtained as above as input parameter and retrieve from MADX the beam position at the two BPMs. %
We then compute the new values of $\delta_p$ using the new values of $y_{BPM_{1,2}}$ and plot on Fig.~\ref{subfig:momentum_offset_error} the difference between the values of $\delta_p$ obtained with $\mu_2 - \mu_1 = 0.5$ and with 
$\mu_2 - \mu_1 = 0.48$, i.e. $\delta_{p_\text{error}}$. %
The distribution shows that the difference between the two $\delta_p$ values is small, most values $<$0.00005 or 2.3\% of the $\delta_p$ distribution RMS value, showing that our $\mu_2 - \mu_1 = 0.5$ approximation is justified.

\section{Betatron contribution}

The betatron oscillation amplitude can be computed using equation~\ref{eq:Main_Assumption} and the momentum offset determined with BPMs 430103 and 430129 (see above). %
These values can then be used to fit equation~\ref{eq:betatron_oscillation_theory} at the position of the BPMs not affected by the presence of the proton bunch. %
The fitted betatron oscillation amplitude can then be extrapolated downstream, to the common BPMs and used to determine the betatron contribution at the BPMs affected by the presence of the proton bunch. %
Positions at these BPMs can then be calculated by also using the momentum offset obtained in the previous paragraph and the dispersion calculated for the beamline. %

\section{Single event results}

\begin{figure}[!htpb]
    \begin{subfigure}[t]{0.49\textwidth}
        \includegraphics[width=\linewidth]{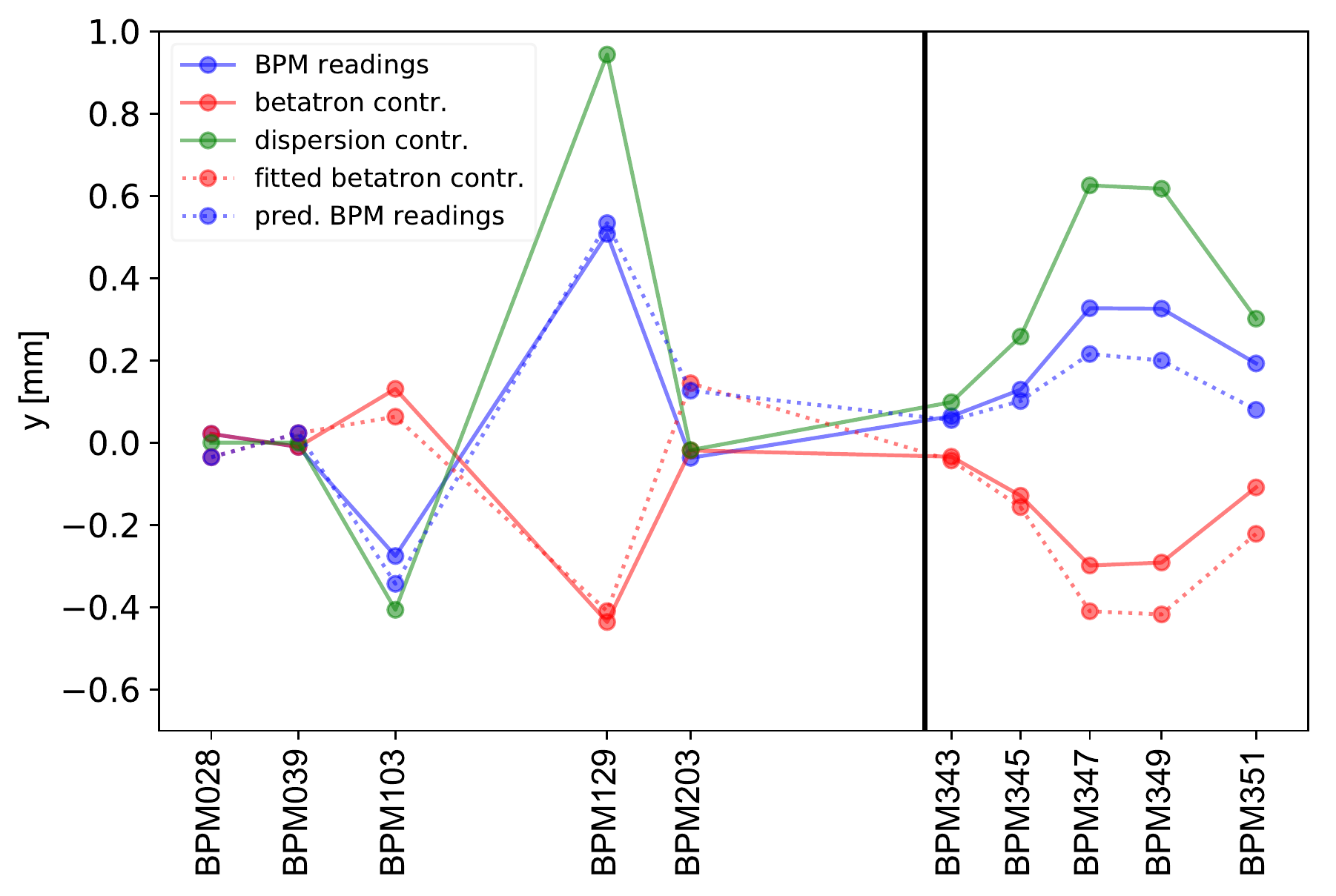}%
        \vspace*{-1.7cm}\caption{\hspace*{5.3cm}\label{subfig:event_5385_y}}
    \end{subfigure}%
    \begin{subfigure}[t]{0.49\textwidth}
        \includegraphics[width=\linewidth]{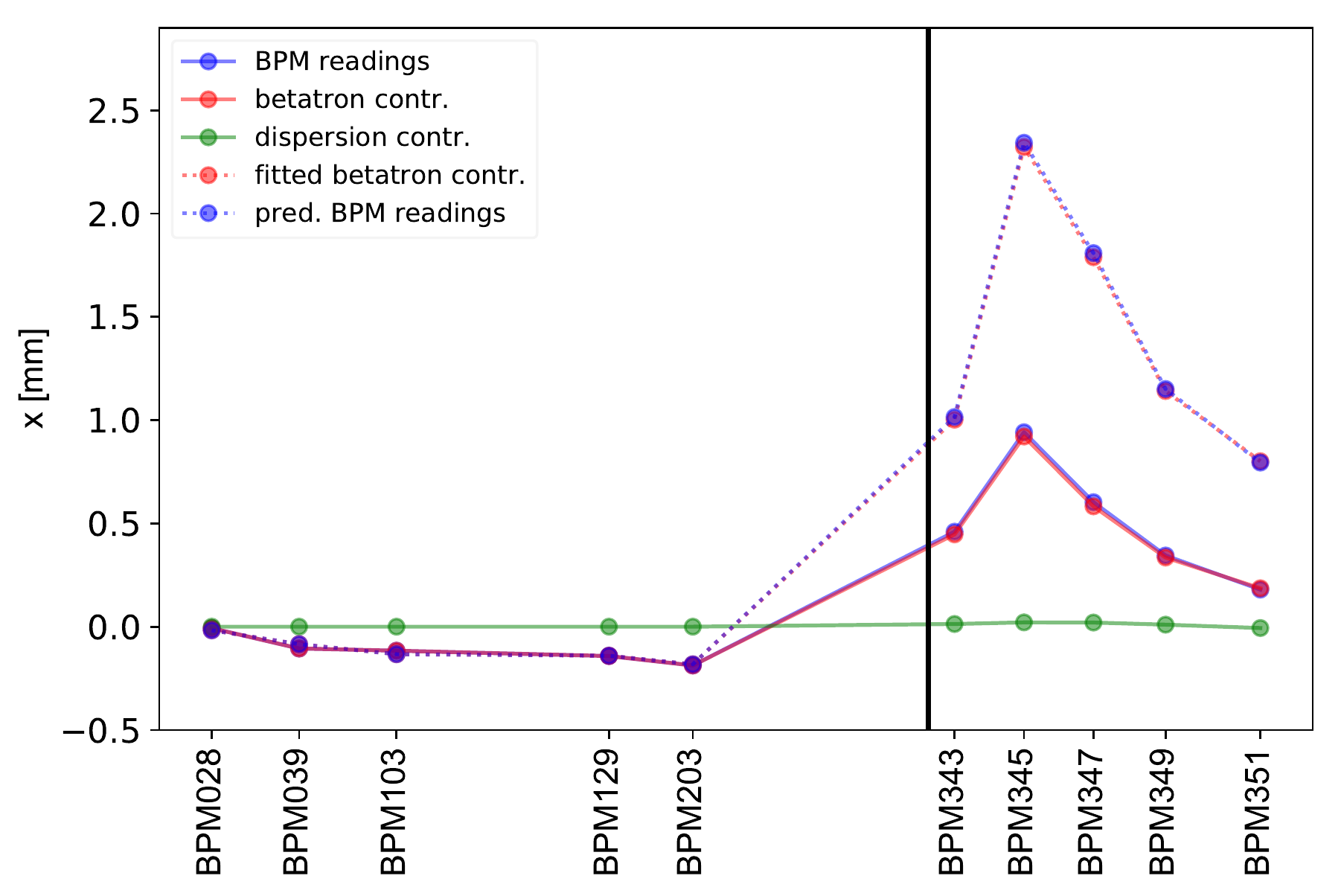}%
        \vspace*{-1.7cm}\caption{\hspace*{5.3cm}\label{subfig:event_5385_x}}
    \end{subfigure}
    \caption{\label{fig:event_5385}Measured transverse position for a single event with respect to the design trajectory in the vertical (a) and horizontal plane (b). %
    Dispersion contribution in green, betatron contribution in red, continuous lines. %
    Betatron contribution based on the fitted betatron oscillation amplitude fitted to BPMs 
    in red, dotted line. 
    Predicted transverse position, i.e. sum of the fitted betatron and dispersion contributions, in blue, dotted line. %
    The vertical black line shows the transition from the electron only beamline (upstream) to the common beamline (downstream). %
    Note in the horizontal plane (b) the overlap of lines due to the negligible dispersion contribution because of negligible dispersion. %
    Also note the different vertical scales.}
\end{figure}

In this case we use the five first BPMs for the  betatron oscillation fit, i.e. with numbers 430028, 430039, 430103, 430129 and 430203. %
The BPM 430308 has proven unreliable during experiments and is therefore not used. %
Further detail can be found in ref.~\cite{bib:felipe}. %
We then compute with the fit the positions at BPMs 412343, 412345, 412347, 412349 and 412351. %
On Fig.~\ref{fig:event_5385} we plot for the x- and y-planes the measured positions (blue symbols, continuous line) and the dispersion contribution (green symbols and continuous line) to the trajectory offset all along the beamline. %
The betatron contribution (red symbols, continuous line) is inferred from the difference between the positions measured and the dispersion contribution. %
We also plot the betatron contribution obtained with the fit of the betatron oscillation amplitude (red symbols and dotted line). %
The predicted trajectory, i.e. the sum of the dispersion and fitted betatron contributions, is shown with blue symbols and dotted line. %
Therefore, one would expect the dotted lines (predicted values) to overlap with the continuous ones (measured values), assuming the dispersion contribution is correct. %
We note that for this event the measured and predicted values in the vertical plane differ by up to 100$ \, \mu$m, but track each other. %
In the horizontal plane the difference is larger (up to $1.5 \,$mm), but again the predicted values track the measured ones. %
The large discrepancy at the end of the line between predicted and measured values could be explained with misalignment of the the strong quadrupole (MQAWF 430311) between the two horizontal dipole magnets used to match the achromatic bend. %

\section{Multiple event results}
\subsection{Prediction with five BPMs}

\begin{figure}[!htpb]
    \begin{subfigure}[t]{0.58\textwidth}
        \includegraphics[width=\linewidth]{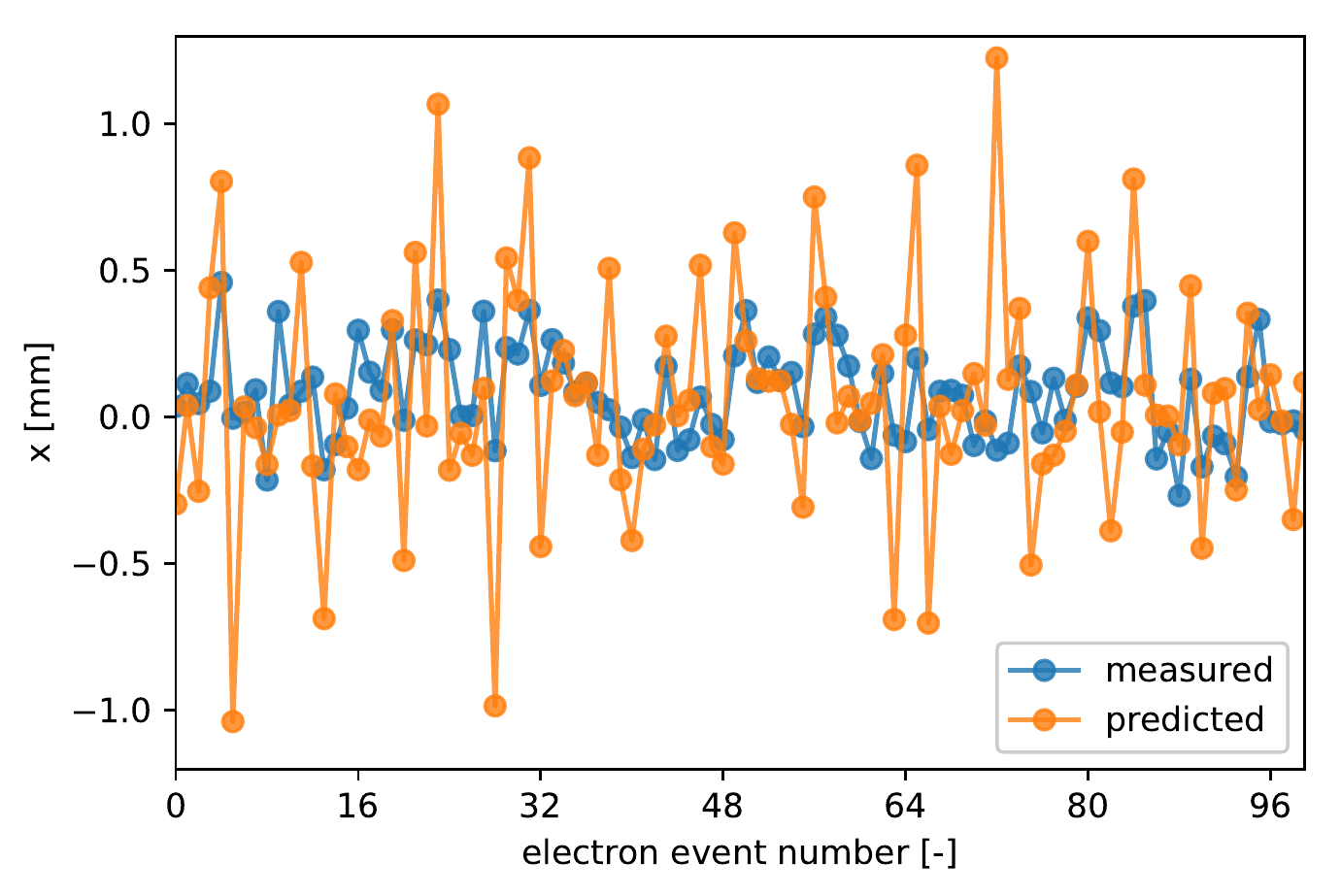}
        \vspace*{-2cm}\caption{\hspace*{5.2cm}\label{subfig:5BPMs_pos_vs_time_x_BPM351}}
    \end{subfigure}%
    \begin{subfigure}[t]{0.42\textwidth}
        \includegraphics[width=\linewidth]{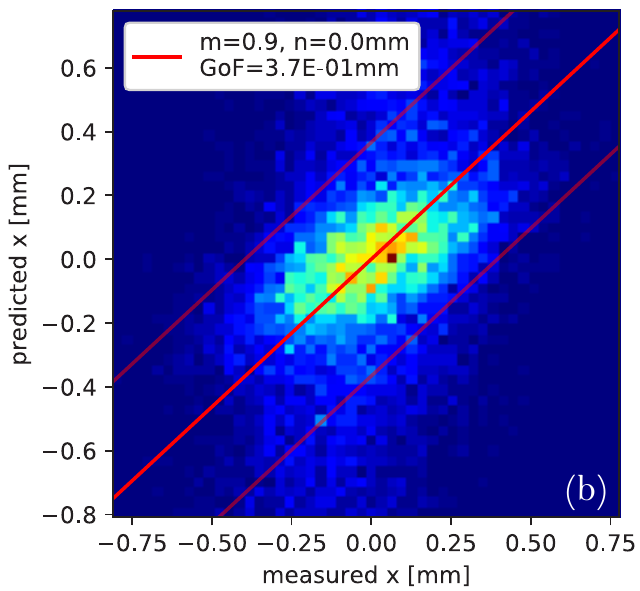}%
        \hypertarget{subfig:5BPMs_correlation_x_BPM351}{}
    \end{subfigure}
    \caption{(a) Measured (blue symbols and line) and predicted (orange symbols and line) values of the beam position in the x-plane at BPM 412351 for 100 consecutive events of the data set. %
    Predictions use the five first BPMs. %
    (b) Two-dimensional density plot (color proportional to number of events from less in blue to more in yellow) showing the correlation between predicted and measured positions for the entire data set. %
    Thick red line: result of linear fit; thin lines, linear fit plus/minus GoF value.}
\end{figure}

Since the purpose of this procedure is to predict the position of the electron beam at the last two BPMs, we show now the measured and predicted values in the horizontal plane at the last BPM 412351 for one hundred consecutive events in Figure~\ref{subfig:5BPMs_pos_vs_time_x_BPM351}. %
The predicted values track the measured values, but their range is larger. %
The distributions of measured and predicted values and their difference for the whole data set ($\num{E4}$ events) are bell-shaped, centered at zero and have a RMS of 220, 420, and $370\,\mu$m, respectively. %

Figure~\hyperlink{subfig:5BPMs_correlation_x_BPM351}{4b} shows a 2D density plot of the predicted versus measured positions for the full set of events. %
It shows that most of the predicted values are within the same range as the measured ones. %
We note that the main distribution is surrounded by a cloud of events with a predicted value range larger than the measured value range. %
This reflects the fact that on Fig.~\ref{subfig:5BPMs_pos_vs_time_x_BPM351} there are two sets of points visible for predicted values: those that are quite close to measured ones, and those that are much farther. %
Since we expect the predicted values to be equal to the measured ones, we use a linear fit between these two sets of points on Fig.~\hyperlink{subfig:5BPMs_correlation_x_BPM351}{4b}. 
The goodness of fit (GoF) is determined as the square root of the mean distance between the predicted values and the linear fit squared, with the parallel lines being vertically shifted by this amount on Fig.~\hyperlink{subfig:5BPMs_correlation_x_BPM351}{4b}. %
The smaller the GoF the stronger measured and predicted values correlate. %
In this case the slope of the linear fit is 0.9, close to the expected value of 1 and the GoF is $370 \, \mu$m.

We obtain similar results in the vertical plane, where the predicted values range is smaller than, but track the measured ones. %
The distributions for measured and predicted values and their difference have a RMS of 190, 60 and 150$ \, \mu$m, respectively. %
The linear fit of the correlation has a slope of 0.3 and a GoF of 31$ \, \mu$m. %

\subsection{Prediction with eight BPMs}
The results above are derived with the five BPMs in the electron beamline. %
Including more BPMs (from the common beamline) can help to validate the model, but cannot be used during acceleration events since their readings are affected by the proton bunch co-propagating with the electron bunch. %
We now show the results obtained when including BPMs 412343, 412345 and 412347, i.e. using eight BPMs for the prediction, for the sake of validation of the procedure. %

\begin{figure}[!htpb]
    \begin{subfigure}[t]{0.58\textwidth}
        \includegraphics[width=\linewidth]{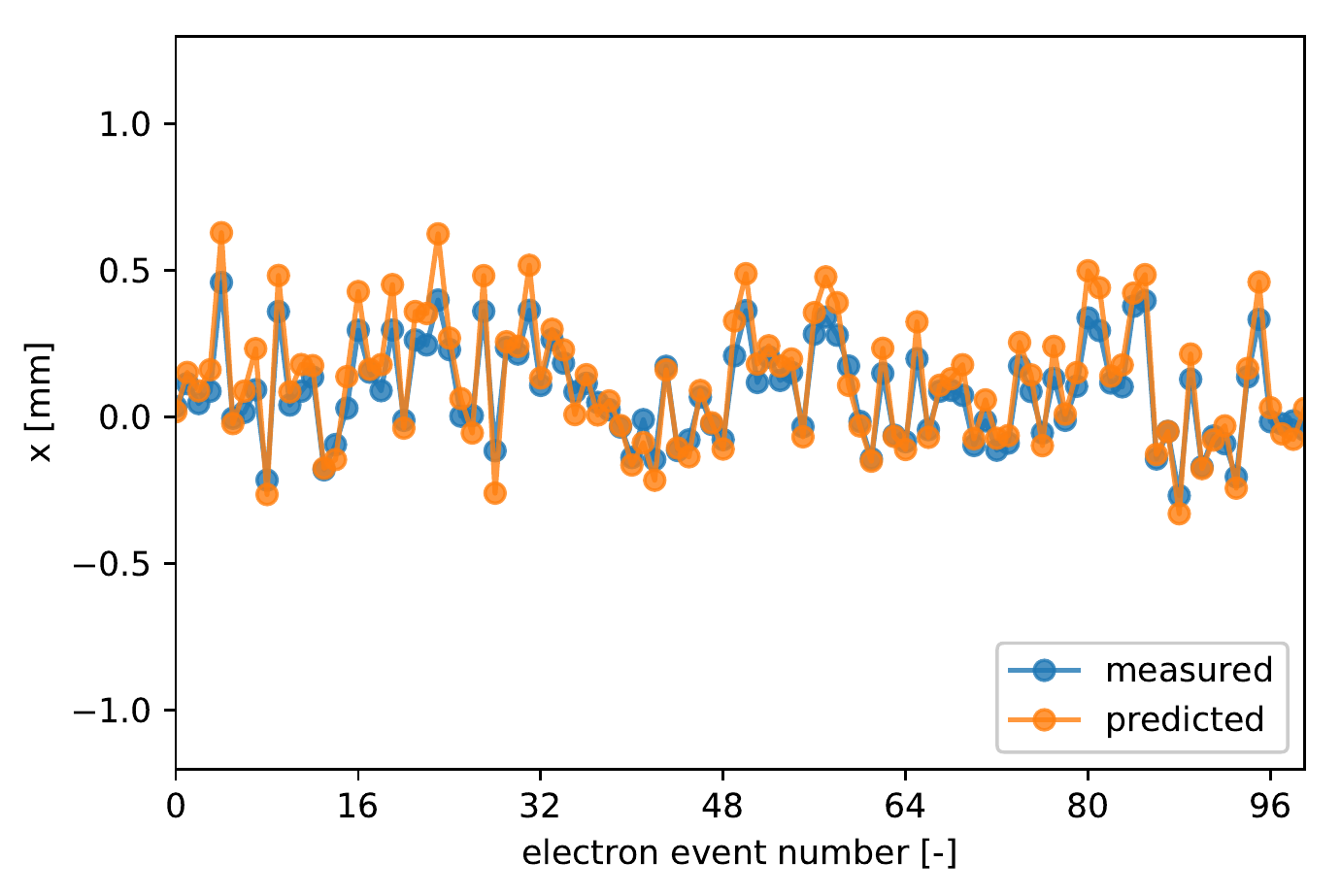}
       \vspace*{-2cm}\caption{\hspace*{6.2cm}\label{subfig:8BPMs_pos_vs_time_x_BPM351}}
    \end{subfigure}%
    \begin{subfigure}[t]{0.42\textwidth}
        \includegraphics[width=\linewidth]{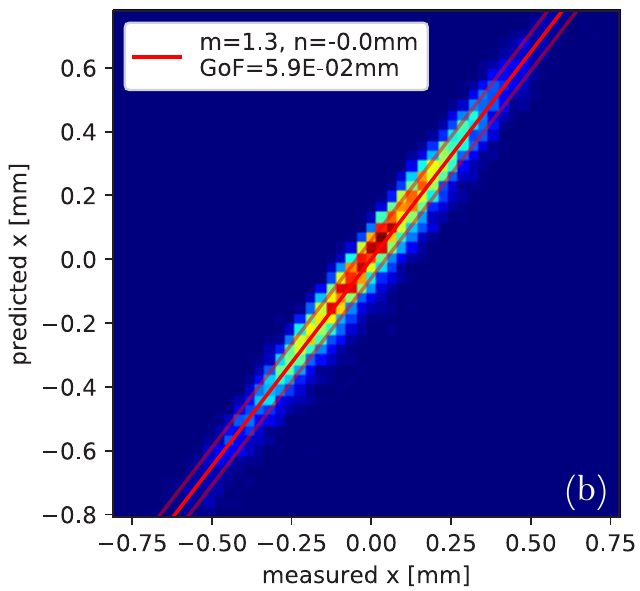}%
        \hypertarget{subfig:8BPMs_correlation_x_BPM351}{}
    \end{subfigure}
    \caption{(a) Measured (blue symbols and line) and predicted (orange symbols and line) values of the beam position in the x-plane at BPM412351 for 100 consecutive events of the data set. %
    Predictions use the eight first BPMs. %
    (b) Two-dimensional density plot (color proportional to number of events from less in blue to more in yellow) showing the correlation between predicted and measured positions for the entire data set. %
    Thick red line: result of linear fit; thin lines, linear fit plus/minus GoF value. }
\end{figure} %

Figure~\ref{subfig:8BPMs_pos_vs_time_x_BPM351} shows that the predicted values with eight BPMs track the measured values better than when using five BPMs (compare with Fig.~\ref{subfig:5BPMs_pos_vs_time_x_BPM351}). %
We see also in Figure~\hyperlink{subfig:8BPMs_correlation_x_BPM351}{5a} a smaller difference between measured and predicted values as the ellipse is more clearly elongated (i.e. smaller GoF, 60 rather than 370$ \, \mu m$). %
The distributions for measured and predicted values and their difference have a RMS of 220, 300, and 90$ \, \mu$m, respectively. %
The difference between the measured and predicted values is now smaller than the range of variations and on the order of the BPM reading resolution~\cite{bib:shengli}. %
This shows that the prediction is much better with eight than with five BPMs. %
The slope of 1.3 (ideally 1.0) shows a systematic error. %

In the vertical plane the predicted values also track better the measured values and the range has increased while still being smaller than the measured range. %
The RMS of the distributions of measured and predicted values and their difference are 190, 100, and 100$ \, \mu$m, respectively. %
The slope of the linear fit of the correlation has increased from 0.3 to 0.5 and the GoF decreased from 31 to 21$ \, \mu$m. %

Similar results were obtained between the measured and predicted values at the previous BPM (number 412349), which would also be used for beams crossing calculations. %

\section{Conclusions}

We use the AWAKE electron beamline model, as well as beam position measurements at five or eight locations in the upstream part of the beamline, to predict the position of the beam in the downstream part of the line. %
Predictions with five BPMs show large variations between predicted and measured positions at the last two BPMs locations, $\sim 370$ in the x-plane and $\sim 150 \, \mu$m in the y-plane for the last BPM. %
For this work we calculated predictions with eight BPMs, which reduced the variations to $\sim 90$ and $\sim 100 \, \mu$m for the horizontal and vertical planes, respectively. %
Similar numbers where obtained for the previous BPM, with $\sim 520$ and $\sim 190 \, \mu$m with five BPMs reduced to $\sim 70$ and $\sim 90 \, \mu$m with eight BPMs for the horizontal and vertical planes, respectively. %
In the real experiments only the first five BPMs can be used since the reading of the last ones is affected by the presence of the proton bunch. %
The geometry of the experiment is such that the electron beam comes from above the proton beam trajectory and is bend down (in the y-plane), towards the proton beam trajectory. %
Errors or variations in the y-plane predictions therefore only affect the z-location of the two beams trajectories crossing, not directly whether they cross or not. %
Errors or variations in the x-plane position directly affect whether the two beams trajectories cross or not.

With a distance of 1.15$\,$m between the last two BPMs and a distance of 2.18\,m to the plasma entrance, this corresponds to an RMS variation in x of about 1.2\,mm, the same order of magnitude as the extent of the plasma radius and larger than the size of the wakefields (0.2\,mm at a plasma electron density of 7$\times$10$^{14}$\,cm$^{-3}$). %
The case is different with eight BPMs, for which the corresponding x-variation is 0.2\,mm. %
The AWAKE collaboration is developing methods e.g.~\cite{bib:Cherenkov} to directly measure the two beams positions at the same time along the common beamline part for each event of future experiments~\cite{bib:muggliRun2}. %

\end{document}